\documentclass{elsarticle}

\usepackage{graphicx}
\usepackage{bm}
\usepackage{amssymb}
\usepackage{amsmath}
\usepackage{amsfonts}

\usepackage[utf8]{inputenc}

\begin{document}

\begin{frontmatter}

\title{$N$-soliton solutions of the Fokas--Lenells
equation for the plasma ion-cyclotron waves: Inverse scattering
transform approach}

\author{V. M. Lashkin\corref{cor}%
\fnref{fn1,fn2}} \ead{vlashkin62@gmail.com}

\cortext[cor]{Corresponding author}

\address[fn1]{Institute for Nuclear Research, Pr. Nauki 47, Kyiv 03028,
Ukraine}
\address[fn2]{Space Research Institute, Pr. Glushkova 40
k.4/1, Kyiv 03187,  Ukraine}

\begin{abstract}
We present a simple and constructive method to find $N$-soliton
solutions of the equation suggested by Davydova and Lashkin to
describe the dynamics of nonlinear ion-cyclotron waves in a plasma
and subsequently known (in a more general form and as applied to
nonlinear optics) as the Fokas--Lenells equation. Using the
classical inverse scattering transform approach, we find bright
$N$-soliton solutions, rational $N$-soliton solutions, and
$N$-soliton solutions in the form of a mixture of exponential and
rational functions. Explicit breather solutions are presented as
examples. Unlike purely algebraic constructions of the Hirota or
Darboux type, we also give a general expression for arbitrary
initial data decaying at infinity, which contains the contribution
of the continuous spectrum (radiation).
\end{abstract}

\begin{keyword}
$N$-soliton solution \sep Fokas--Lenells equation \sep
exponential-rational solution \sep inverse scattering transform
\sep continuous spectrum \sep ion-cyclotron waves
\end{keyword}
\end{frontmatter}

\section{Introduction}

Equations integrable by the inverse scattering transform (IST)
method, which arise in real physical situations and are important
for practical applications, are of particular interest in
nonlinear science \cite{Scott_nonlin}. Generally speaking, the
number of such equations is very limited and not too large. In
plasma physics, classical examples of completely integrable
equations are the Korteweg-de Vries (KdV) equation (and the
modified KdV equation) for the nonlinear ion-acoustic waves, the
nonlinear Schr\"{o}dinger (NLS) equation for the Langmuir waves
(both of these equations are also derived for the cases of other
branches of plasma oscillations using the reductive perturbation
technique), the derivative nonlinear Schr\"{o}dinger (DNLS)
equation describing nonlinear Alfv\'{e}n waves, and the
two-dimensional Kadomtsev-Petviashvili equation, which is a
two-dimensional generalization of the KdV equation
\cite{Petviashvili_book1992,Scorich2010}. Slightly less known
integrable models (in plasmas) are the Boussinesq equation for the
beam instabilities \cite{Kono} and the nonlinear string equation
(elliptic Boussinesq) describing a nonlinear stage of the
two-stream instability in quantum plasmas \cite{Lashkin_string},
and also the Yajima-Oikawa equations describing the interaction of
Langmuir waves with ion-acoustic waves propagating in one
direction \cite{Oikawa}.

Davydova and Lashkin \cite{Lashkin1991} suggested a nonlinear
equation governing the dynamics of short-wavelength ion-cyclotron
waves in plasmas (the Bernstein modes)
\cite{Krall_book1973,Akhiezer_book1975}, which in the
one-dimensional case in dimensionless variables has the form
\begin{equation}
\label{main} u_{xt}-u-i\sigma|u|^{2}u_{x}=0,
\end{equation}
where $u(x,t)$ is the slowly varying complex envelope of the
electrostatic potential at the ion-cyclotron frequency, and
$\sigma=\pm 1$. Authors of \cite{Lashkin1991} found the bright
one-soliton solution of (\ref{main}), and then the same authors
(co-authored with A. I. Fishchuck) presented solutions of this
equation in the form of bright algebraic soliton, dark and
anti-dark solitons corresponding the nonvanishing boundary
conditions, as well as solutions in the form of nonlinear periodic
waves in elliptic Jacobi functions \cite{Lashkin1994}. Later on,
Fokas and Lenells showed \cite{Fokas1995,Lenells2009_Nonlinearity}
that (\ref{main}) is completely integrable and corresponds to the
first negative flow of the Kaup--Newell hierarchy of the DNLS
equation \cite{Kaup1978}, and, therefore, can be solved by the IST
\cite{Zakharov_book}. Note that the original version of the
equation considered in \cite{Fokas1995,Lenells2009_Nonlinearity}
differs from (\ref{main}) and has been derived as an integrable
generalization of the NLS equation using bi-Hamiltonian methods
\cite{Fokas1995}, and then as a model for nonlinear pulse
propagation in monomode optical fibers when certain higher-order
nonlinear effects are taken into account
\cite{Lenells2009_derivation}. Under this, the corresponding
equation in dimensionless variables is
\begin{equation}
\label{Fokas_Lenells1} iu_{t}-\nu u_{xt}+\gamma u_{xx}+\sigma
|u|^{2} (u+i\nu u_{x})=0,
\end{equation}
where $\nu$ and $\gamma$ are real constants, $\sigma=\pm 1$, and
by gauge transformation and a change of variables can be reduced
to equation (\ref{main}) suggested in \cite{Lashkin1991}. Lenells
rediscovered \cite{Lenells2009_derivation} the bright one-soliton
solution of \cite{Lashkin1991} without using the IST. Bright
$N$-soliton solutions of (\ref{Fokas_Lenells1}) were obtained by
Lenells in \cite{Lenells_N-soliton2010} with the dressing method,
and for (\ref{main}) by the Hirota bilinear method in
\cite{Matsuno_bright2012}. Dark $N$-soliton solutions, which
contain dark and anti-dark soliton solutions of
\cite{Lashkin1994}, were found by the bilinearization method in
\cite{Veksler_dark2011,Matsuno_dark2012}. The $N$-order rogue wave
solution of equations (\ref{main}) and (\ref{Fokas_Lenells1}) were
obtained using the $N$-fold Darboux transformation in
\cite{Geng2019,Xu2015}. In what follows, we will refer to
(\ref{main}) as the Davydova-Lashkin-Fokas-Lenells (DLFL)
equation. The DLFL equation (\ref{main}) is universal in the sense
that it contains only three terms of the second of which
corresponds to weak dispersion ($\omega\sim 1/k\ll 1$) and the
third to weak (cubic) nonlinearity. Here, $\omega$ and $k$ are the
frequency and wave number respectively, where in the linear part
$u\sim \exp (i\omega t-ikx)$. The same situation holds for  the
NLS and DNLS equations with the weak dispersion $\omega\sim
k^{2}\ll 1$ (and cubic nonlinearity), and the KdV equation with
$\omega\sim k^{3}\ll 1$ (and quadratic nonlinearity). Note that
the weak dispersion and nonlinearity in all these cases follow
from the physical derivation of the corresponding equations
\cite{Petviashvili_book1992}.

Unlike equation (\ref{Fokas_Lenells1}) derived for a very special
case of pulse propagation in nonlinear optics, the DLFL equation
(\ref{main}) describes nonlinear one-dimensional short-wavelength
ion-cyclotron waves in plasmas. The importance of the theoretical
study of nonlinear ion-cyclotron waves, in particular solitons, is
due, first of all, to reliable experimental data on their
observation in the Earth's magnetosphere
\cite{Andre1987,Cattell1991,Temerin1997}. In particular, the
profile of the measured electric field is definitely a chain of
several solitons \cite{Temerin1997}.

The aim of this paper is to obtain the $N$-soliton solutions of
the DLFL equation using the classical method of the IST in which
it is possible to take into account the continuous spectrum of the
spectral problem (radiation). Note that purely algebraic methods
for finding the $N$-soliton solutions such as the Darboux
transformation or the Hirota bilinear method are not suitable for
this purpose by definition. Note that in practical situations, the
initial perturbation almost never corresponds to a purely
solitonic (reflectionless potential) and can be either quite close
or very different from it. In any case, then taking into account
the continuous spectrum and using IST (in one form or another) is
necessary. From a physical point of view, the spectral parameter
of the continuous spectrum $\lambda$ in the IST for the DLFL
equation is related to the wave number of emitted quasilinear
ion-cyclotron waves $k$ by a simple relation \cite{Lashkin2021}.
We use the IST in its classical form, but all results can be
easily reformulated and reproduced using the Riemann-Hilbert
problem.

The paper is organized as follows. In section \ref{Sec2} we review
some results on the IST for the DLFL equation, and then present
the general formal solution as the sum of the soliton and
nonsoliton parts expressed in terms of the scattering data and
Jost solutions corresponding to the discrete and continuous
spectrum, respectively. In section \ref{Sec3} we find the
$N$-soliton solutions and, in particular, algebraic $N$-soliton
solutions and solutions in the form of a mixture of rational and
exponential functions. The asymptotics of the $N$-soliton solution
is considered in section \ref{Sec4}. The conclusion is made in
section \ref{Sec5}. In the Appendix, we give a brief outline of
the derivation of a two-dimensional nonlinear equation describing
the dynamics of ion-cyclotron waves in plasmas, which in the
one-dimensional case reduces to the DLFL equation (\ref{main}).

\section{\label {Sec2} Spectral problem for the DLFL equation}

At the beginning of this section we will give a brief overview on
the IST for the DLFL equation following \cite{Lashkin2021}. The
DLFL equation (\ref{main}) can be written as the compatibility
condition
\begin{equation}
\label{compatib}
\mathbf{U}_{t}-\mathbf{V}_{x}+[\mathbf{U},\mathbf{V}]=0 ,
\end{equation}
of two linear matrix equations \cite{Lenells2009_Nonlinearity,
Lashkin2021}
\begin{eqnarray}
\label{spec1} \mathbf{M}_{x}=\mathbf{U}\mathbf{M}, \\
\label{spec2} \mathbf{M}_{t}=\mathbf{V}\mathbf{M},
\end{eqnarray}
where
\begin{eqnarray}
\label{U} \mathbf{U}=-i\lambda^{2}\sigma_{3}+\lambda
\mathbf{Q}_{x}, \,\,\,\, \mathbf{Q}=\left(\begin{array}{cc} 0 & u
\\ \sigma u^{\ast} & 0
\end{array}\right)
\\
\label{V}
\mathbf{V}=\frac{i}{4\lambda^{2}}\sigma_{3}-\frac{i}{2\lambda}\sigma_{3}\mathbf{Q}
+\frac{i}{2}\sigma_{3}\mathbf{Q}^{2},
\end{eqnarray}
and where $\mathbf{M}(x,t,\lambda)$ is a $2\times 2$ matrix-valued
function, $\lambda$ is a complex spectral parameter and
$\sigma_{3}$ is the Pauli matrix. The Jost solutions
$\mathbf{M}^{\pm}(x,t,\lambda)$ of  (\ref{spec1}) for real
$\lambda^{2}$ and for some fixed $t$ ($t$-dependence will be
omitted for now) are defined by the boundary conditions
\begin{equation}
\label{boundary} \mathbf{M}^{\pm}(x,\lambda)\rightarrow \exp
(-i\lambda^{2}\sigma_{3}x)
\end{equation}
as $x\rightarrow\pm \infty$. The matrix Jost solutions
$\mathbf{M}^{\pm}$ are presented in the form
\begin{equation}
\label{M-plus-minis}
\mathbf{M}^{+}(x,\lambda)=\left(\begin{array}{cc}
\tilde{\psi}_{1} (x,\lambda) & \psi_{1} (x,\lambda) \\
\tilde{\psi}_{2} (x,\lambda) & \psi_{2} (x,\lambda)
\end{array}\right), \,\,
\mathbf{M}^{-}(x,\lambda)=\left(\begin{array}{cc}
\varphi_{1} (x,\lambda) & -\tilde{\varphi}_{1}(x,\lambda) \\
\varphi_{2} (x,\lambda) & -\tilde{\varphi}_{2} (x,\lambda)
\end{array}\right).
\end{equation}
The scattering matrix $\mathbf{S}$
\begin{equation}
\renewcommand*{\arraystretch}{1.3}
\mathbf{S}(\lambda)= \left(\begin{array}{cc}
 a(\lambda)  & -\tilde{b}(\lambda) \\
 b(\lambda) & \tilde{a}(\lambda)
\end{array}\right)
\end{equation}
with $a\tilde{a}+b\tilde{b}=1$ relates the two fundamental
solutions $\mathbf{M}^{-}$ and $\mathbf{M}^{+}$
\begin{equation}
\label{Scat_matrix}
\mathbf{M}^{-}(x,\lambda)=\mathbf{M}^{+}(x,\lambda)\mathbf{S}(\lambda),
\end{equation}
so that
\begin{eqnarray}
\varphi=a\tilde{\psi}+b\psi,
\label{co1} \\
\tilde{\varphi}=-\tilde{a}\psi+\tilde{b}\tilde{\psi}.
 \label{co2}
\end{eqnarray}
Further we set $\sigma=-1$ without loss of generality. It follows
from  (\ref{spec1}) and (\ref{Scat_matrix}) that matrices
$\mathbf{M}^{\pm}$ and $\mathbf{S}$ have the parity symmetry
properties,
\begin{equation}
\label{parity}
\mathbf{M}^{\pm}(x,\lambda)=\sigma_{3}\mathbf{M}^{\pm}(x,-\lambda)\sigma_{3},
\,\, \mathbf{S}(\lambda)=\sigma_{3}\mathbf{S}(-\lambda)\sigma_{3},
\end{equation}
and the conjugation symmetry properties
\begin{equation}
\label{conjugation}
\mathbf{M}^{\pm}(x,\lambda)=\sigma_{2}\mathbf{M}^{\pm
\ast}(x,\lambda^{\ast})\sigma_{2},
\end{equation}
\begin{equation}
\label{conjugation_a_b}
\tilde{a}(\lambda)=a^{\ast}(\lambda^{\ast}), \quad \,
\tilde{b}(\lambda)= b^{\ast}(\lambda^{\ast}),
\end{equation}
where $\sigma_{2}$ and $\sigma_{3}$ are Pauli matrices. The
coefficients $a(\lambda)$ and $b(\lambda)$ are
\begin{equation}
\label{S11} a(\lambda)=\mathrm{det}\,(\varphi,\psi), \,\,\,
b(\lambda)=\mathrm{det}\,(\tilde{\psi},\varphi).
\end{equation}
The zeros $\lambda_{j}^{2}$ ($j=1\dots N$) of the function
$a(\lambda)$ in the region of its analiticity
$\mathrm{Im}\,\lambda^{2}>0$ (correspondingly, the zeros
$\lambda_{j}^{\ast 2}$ of the function $\tilde{a}(\lambda)$ in the
region $\mathrm{Im}\,\lambda^{2}<0$ ) give the discrete spectrum
of the linear problem (\ref{spec1}) and correspond to solitons.
The zeros $\lambda_{j}$ ($j=1\dots 2N$) appear in pairs and one
can choose $\lambda_{j}$ ($j=1\dots N$) in the first quadrant and
$\lambda_{j+N}=-\lambda_{j}$ in the third quadrant. Then, as it
follows from (\ref{co1}) and (\ref{co2}), the functions
$\varphi(x,\lambda_{j})$ and $\psi(x,\lambda_{j})$ are linearly
dependent
\begin{equation}
\label{bj}
\varphi(x,\lambda_{j})=b_{j}(\lambda_{j})\psi(x,\lambda_{j}),\quad
\tilde{\varphi}(x,\lambda_{j}^{\ast})=
b_{j}^{\ast}(\lambda_{j}^{\ast})\tilde{\psi}(x,
\lambda_{j}^{\ast}).
\end{equation}
The coefficient $a(\lambda)$ can be expressed in terms of its
zeros and the values of $b(\lambda)$ on the contour
$\Gamma=(+\infty,0)\bigcup (-\infty,0)\bigcup (+i\infty,0)\bigcup
(-i\infty,0)$ \cite{Lashkin2021},
\begin{equation}
a(\lambda)=\prod_{j=1}^{N}\frac{\lambda_{j}^{\ast 2
}}{\lambda_{j}^{2}}\frac{(\lambda^{2}-\lambda_{j}^{2})}
{(\lambda^{2}-\lambda_{j}^{\ast 2 })}\exp \left\{\frac{1}{2\pi i}
\int_{\Gamma}\frac{\lambda^{2}\ln(1+\sigma\,\mathrm{sgn}\,\mu^{2}|\,b(\mu)|^{\,2})}{\mu(\mu^{2}-
\lambda^{2})}\,d\mu\right\}. \label{mnls_reflection1}
\end{equation}
An important particular case is that of the solitonic
("reflectionless") potentials $u(x)$ when $b(\lambda,t)=0$ as a
function of $\lambda$ for some fixed $t$. It then follows from
(\ref{mnls_reflection1}) that
\begin{equation}
\label{a} a(\lambda)=\prod_{j=1}^{N}\frac{\lambda^{\ast
2}_{j}}{\lambda^{2}_{j}}
\frac{(\lambda^{2}-\lambda^{2}_{j})}{(\lambda^{2}-\lambda^{\ast 2
}_{j})}.
\end{equation}
The  time evolution of the scattering data, as usual in the IST,
turns out to be trivial,
\begin{eqnarray}
\label{dyn1} \lambda_{j}(t)=\lambda_{j}(0),
\\
\label{dyn2} b_{j}(t)=b_{j}(0)\exp[-i/(2\lambda^{2}_{j})t],
\\
\label{dyn3} b(\lambda,t)=b(\lambda,0)\exp[-i/(2\lambda^{2})t].
\end{eqnarray}
and in the following we denote $\lambda_{j}(t)\equiv \lambda_{j}$,
$b_{j}(t)\equiv b_{j}$ and $b(\lambda,t)\equiv b(\lambda)$. Taking
into account the boundary conditions (\ref{boundary}), the
corresponding integral equations for $\mathbf{M}^{+}$ can be
obtained from (\ref{spec1}):
\begin{eqnarray}
\label{psi1}
\psi_{1}(x,\lambda)=-\lambda\int_{x}^{\infty}\mathrm{e}^{-i\lambda^{2}(x-y)}
u_{y}\psi_{2}(y,\lambda)\,dy,
\\
\label{psi2}
\psi_{2}(x,\lambda)=\mathrm{e}^{i\lambda^{2}x}+\lambda\int_{x}^{\infty}
\mathrm{e}^{i\lambda^{2}(x-y)}u_{y}^{\ast}\psi_{1}(y,\lambda)\,dy,
\\
\label{psi-tild1}
\tilde{\psi}_{1}(x,\lambda)=\mathrm{e}^{-i\lambda^{2}x}-\lambda\int_{x}^{\infty}
\mathrm{e}^{-i\lambda^{2}(x-y)}u_{y}\tilde{\psi}_{2}(y,\lambda)\,dy,
\\
\label{psi-tild2}
\tilde{\psi}_{2}(x,\lambda)=\lambda\int_{x}^{\infty}
\mathrm{e}^{i\lambda^{2}(x-y)}u_{y}^{\ast}\tilde{\psi}_{1}(y,\lambda)\,dy.
\end{eqnarray}
Then, from (\ref{psi1})-(\ref{psi-tild2}) one can find the
corresponding asymptotics at $\lambda\rightarrow 0$,
\begin{eqnarray}
\label{ps1} \psi_{1}(x,\lambda)=\lambda u+O(\lambda^{2}),
\\
\label{ps2} \psi_{2}(x,\lambda)=1+O(\lambda^{2}),
\\
\label{ps-tild1} \tilde{\psi}_{1}(x,\lambda)=1+O(\lambda^{2}),
\\
\label{ps-tild2} \tilde{\psi}_{2}(x,\lambda)=-\lambda
u^{\ast}+O(\lambda^{2}).
\end{eqnarray}
Note that equation (\ref{spec1}), that is the $x$ part of the Lax
pair (\ref{spec1})-(\ref{spec2}) of the DLFL equation, is simply
related to the $x$ part of the Lax pair of the DNLS equation by
the replacement $u \rightarrow u_{x}$. The revised Zakharov
equations for the Jost functions $\psi$ and $\tilde{\psi}$ of the
DNLS equation were obtained in \cite{Huang2007} and coincide with
the corresponding equations of the DLFL equation, except that the
time dependences of the coefficients $b_{j}(t)$ and $b(\lambda,t)$
are determined by equations (\ref{dyn2}) and (\ref{dyn3})
respectively. Following \cite{Huang2007} and using this analogy,
we can write the equations for  $\tilde{\psi}_{1,2}$ in the form
\begin{eqnarray}
\label{Ps1}
\tilde{\psi}_{1}(x,\lambda)=\mathrm{e}^{-i\lambda^{2}x}
+\sum_{k=1}^{2N}\frac{\lambda^{2}}{\lambda_{k}^{2}(\lambda-\lambda_{k})}
\frac{b_{k}(\lambda_{k})}{\dot{a}(\lambda_{k})}
\psi_{1}(x,\lambda_{k})\mathrm{e}^{i(\lambda^{2}_{k}-\lambda^{2})x}
\nonumber \\
+\frac{1}{2\pi i}\int_{\Gamma}\frac{\lambda^{2}}{\lambda^{'
2}(\lambda^{'
}-\lambda)}\frac{b(\lambda^{'})}{a(\lambda^{'})}\psi_{1}(x,\lambda^{'})\mathrm{e}^{i(\lambda^{'
2}-\lambda^{2})x}\,d\lambda^{'},
\\
\label{Ps2}
\tilde{\psi}_{2}(x,\lambda)=\sum_{k=1}^{2N}\frac{\lambda}{\lambda_{k}(\lambda-\lambda_{k})}
\frac{b_{k}(\lambda_{k})}{\dot{a}(\lambda_{k})}
\psi_{2}(x,\lambda_{k})\mathrm{e}^{i(\lambda^{2}_{k}-\lambda^{2})x}
\nonumber \\
+\frac{1}{2\pi i}\int_{\Gamma}\frac{\lambda}{\lambda^{'
}(\lambda^{'
}-\lambda)}\frac{b(\lambda^{'})}{a(\lambda^{'})}\psi_{2}(x,\lambda^{'})\mathrm{e}^{i(\lambda^{'
2}-\lambda^{2})x}\,d\lambda^{'}.
\end{eqnarray}
Then from (\ref{ps-tild2}) and (\ref{Ps2}) we have
\begin{equation}
\label{psi-0} u^{\ast}=-\lim_{\lambda\rightarrow
0}\frac{\tilde{\psi}_{2}(x,\lambda)}{\lambda}=u^{\ast}_{s}+u^{\ast}_{rad},
\end{equation}
where $u^{\ast}_{s}$ corresponds to the discrete part of the
spectrum (solitons),
\begin{equation}
\label{u_sol} u^{\ast}_{s}=\sum_{k=1}^{N}\frac{2}{\lambda_{k}^{2}}
\frac{b_{k}(\lambda_{k})}{\dot{a}(\lambda_{k})}
\psi_{2}(x,\lambda_{k})\mathrm{e}^{i\lambda_{k}^{2}x},
\end{equation}
and we have taken into account the reduction properties
(\ref{parity}) and (\ref{conjugation}), using which the sum over
$2N$ terms in (\ref{Ps2}) is replaced by the sum over $N$. The
term $u^{\ast}_{rad}$ corresponds to the continuous spectrum
(radiation field),
\begin{equation}
\label{u_rad} u^{\ast}_{rad}=\frac{1}{2\pi
i}\int_{\Gamma}\frac{\psi_{2}(x,\lambda)}{\lambda^{2}}\frac{b(\lambda)}{a(\lambda)}
\mathrm{e}^{i\lambda^{2}x}\,d\lambda.
\end{equation}
In the general case, as in other integrable models such as the NLS
equation, KdV equation, etc., an arbitrary initial perturbation
vanishing at infinity rapidly enough decays over time into
dispersive quasilinear waves corresponding to $u_{rad}$ and
solitons corresponding to $u_{s}$ (if any - depending on the
initial conditions, the solitons may not occur at all). In
contrast to the work of Matsuno \cite{Matsuno_bright2012}, where
the purely algebraic Hirota bilinear method was used to find
$N$-soliton solutions, as well as the work of Lenells
\cite{Lenells_N-soliton2010} using the dressing method, the
expression (\ref{psi-0}) also contains the nonsoliton part
$u_{rad}$ associated with the nonzero coefficient $b(\lambda)$
(or, equivalently, to $r(\lambda)=b(\lambda)/a(\lambda)$ sometimes
called the reflection coefficient in the IST). The initial
conditions corresponding to purely $N$-soliton solutions, as is
known, correspond to $b(\lambda)=0$ (the so-called reflectionless
potentials). As was shown in \cite{Lashkin2021}, considering the
radiative component as a superposition of free waves governed by
the linearized equation (\ref{main}) with the dispersion law
$\omega=1/k$, one can conclude that the spectral parameter
$\lambda$ is connected to the wave number of the emitted
quasilinear waves $k$ by the relation
\begin{equation}
\label{k-rel} k=2\lambda^{2}.
\end{equation}
In the general case, when $b(\lambda)\neq0$, as is known, the
solution cannot be written in an explicit analytical form,
however, if $b(\lambda)$ is not equal to zero, but small enough,
$b(\lambda)\ll 1$, one can use the perturbation theory. Under
this, the coefficient $a(\lambda)$ and the Jost function
$\psi_{2}(x,\lambda)$ can be taken purely soliton (the calculation
of the purely $N$-soliton $\psi_{2}(x,\lambda)$ is described in
the next section). Continuous spectrum effects, in particular, the
spectral distribution of ion-cyclotron wave radiation within the
framework of the DLFL equation (\ref{main}), were considered in
\cite{Lashkin2021}. An example of calculating of the radiation
field $u_{rad}$ in the physical space for the DNLS equation is
given in \cite{Lashkin2006}.

\section{\label {Sec3} $N$-soliton solutions}

In the pure soliton case $b(\lambda)=0$, using the parity and
conjugation properties (\ref{parity}) and (\ref{conjugation}) in
equations (\ref{Ps1}) and (\ref{Ps2}), we have for the Jost
solutions $\tilde{\psi}_{1}(x,\lambda)$ and $\tilde{\psi}_{2}$,
\begin{eqnarray}
\label{Psi1}
\tilde{\psi}_{1}(x,\lambda)=\mathrm{e}^{-i\lambda^{2}x}
+\sum_{k=1}^{N}\frac{2\lambda^{2}}{\lambda_{k}(\lambda^{2}-\lambda^{2}_{k})}
\frac{b_{k}(\lambda_{k})}{\dot{a}(\lambda_{k})}
\psi_{1}(x,\lambda_{k})\mathrm{e}^{i(\lambda^{2}_{k}-\lambda^{2})x}
\\
\label{Psi2}
\tilde{\psi}_{2}(x,\lambda)=\sum_{k=1}^{N}\frac{2\lambda}{(\lambda^{2}-\lambda^{2}_{k})}
\frac{b_{k}(\lambda_{k})}{\dot{a}(\lambda_{k})}
\psi_{2}(x,\lambda_{k})\mathrm{e}^{i(\lambda^{2}_{k}-\lambda^{2})x}.
\end{eqnarray}
Evaluating (\ref{Psi1}) and (\ref{Psi2}) at $\lambda_{j}^{\ast}$
and taking into account the conjugation properties
(\ref{conjugation}), one can obtain
\begin{eqnarray}
\label{Psi11} \psi_{2
}^{\ast}(x,\lambda_{j})=\mathrm{e}^{-i\lambda^{\ast 2}_{j}x}
+\sum_{k=1}^{N}\frac{2\lambda^{\ast
2}_{j}}{\lambda_{k}(\lambda^{\ast 2}_{j}-\lambda^{2}_{k})}
\frac{b_{k}(\lambda_{k})}{\dot{a}(\lambda_{k})}
\psi_{1}(x,\lambda_{k})\mathrm{e}^{i(\lambda^{2}_{k}-\lambda^{\ast
2}_{j})x}
\\
\label{Psi22}
\psi_{1}^{\ast}(x,\lambda_{j})=-\sum_{k=1}^{N}\frac{2\lambda_{j}^{\ast}}
{(\lambda^{\ast 2}_{j}-\lambda^{2}_{k})}
\frac{b_{k}(\lambda_{k})}{\dot{a}(\lambda_{k})}
\psi_{2}(x,\lambda_{k})\mathrm{e}^{i(\lambda^{2}_{k}-\lambda^{\ast
2}_{j})x}.
\end{eqnarray}
Equations (\ref{Psi11}) (after its complex conjugation) and
(\ref{Psi22}) are a system of $2N$ linear algebraic equations for
the vector functions $\psi_{1}^{\ast}(x,\lambda_{j})$ and
$\psi_{2}(x,\lambda_{j})$. This system can be solved in a standard
way, and after obtaining $\psi_{2}(x,\lambda_{j})$ and using
(\ref{u_sol}) one can find a solution $u^{\ast}$ through the
corresponding determinants. A similar procedure was used to obtain
$N$-soliton solutions of the DNLS equation \cite{Huang2007}. Here,
however, we present a simple alternative way of finding
$\psi_{2}(x,\lambda_{j})$, which leads to a much more compact
formula for the $N$-soliton solution of equation (\ref{main}). An
analogue of equation (\ref{Psi1}) for the function
$\varphi_{1}(x,\lambda)$ can be written in the form
\begin{equation}
\label{Phi1} \varphi_{1}(x,\lambda)=\mathrm{e}^{-i\lambda^{2}x}
+\sum_{k=1}^{N}\frac{2\lambda^{2}}{\lambda_{k}^{\ast}(\lambda^{2}-\lambda^{\ast
2}_{k})}
\frac{b_{k}(\lambda_{k}^{\ast})}{\dot{\tilde{a}}(\lambda_{k}^{\ast})}
\tilde{\varphi}_{1}(x,\lambda_{k}^{\ast})\mathrm{e}^{i(\lambda^{\ast
2}_{k}-\lambda^{2})x}.
\end{equation}
Evaluating this at $\lambda_{j}$ we have
\begin{equation}
\label{Phi-1}
\varphi_{1}(x,\lambda_{j})=\mathrm{e}^{-i\lambda^{2}_{j}x}
+\sum_{k=1}^{N}\frac{2\lambda^{2}_{j}}{\lambda_{k}^{\ast}(\lambda^{2}_{j}-\lambda^{\ast
2}_{k})}
\frac{b_{k}(\lambda_{k}^{\ast})}{\dot{\tilde{a}}(\lambda_{k}^{\ast})}
\tilde{\varphi}_{1}(x,\lambda_{k}^{\ast})\mathrm{e}^{i(\lambda^{\ast
2}_{k}-\lambda^{2}_{j})x}.
\end{equation}
Using
$\dot{\tilde{a}}(\lambda_{k}^{\ast})=\dot{a}^{\ast}(\lambda_{k})$
and $b_{k}(\lambda_{k}^{\ast})b_{k}^{\ast}(\lambda_{k})=1$ we find
\begin{equation}
\label{Phi11}
b_{j}(\lambda_{j})\psi_{1}(x,\lambda_{j})=\mathrm{e}^{-i\lambda^{2}_{j}x}
+\sum_{k=1}^{N}\frac{2\lambda^{2}_{j}}{\lambda_{k}^{\ast}(\lambda^{2}_{j}-\lambda^{\ast
2}_{k})} \frac{1}{\dot{a}^{\ast}(\lambda_{k})}
\psi_{2}^{\ast}(x,\lambda_{k})\mathrm{e}^{i(\lambda^{\ast
2}_{k}-\lambda^{2}_{j})x}.
\end{equation}
On the other hand, taking complex conjugate of (\ref{Psi22}) and
then multiplying it by $b_{j}(\lambda_{j})$ we have
\begin{equation}
\label{Psi222}
b_{j}(\lambda_{j})\psi_{1}(x,\lambda_{j})=-\sum_{k=1}^{N}\frac{2\lambda_{j}b_{j}(\lambda_{j})}
{(\lambda^{2}_{j}-\lambda^{\ast 2}_{k})}
\frac{b_{k}^{\ast}(\lambda_{k})}{\dot{a}^{\ast}(\lambda_{k})}
\psi_{2}^{\ast}(x,\lambda_{k})\mathrm{e}^{-i(\lambda^{\ast
2}_{k}-\lambda^{2}_{j})x}.
\end{equation}
Subtracting (\ref{Psi222}) from (\ref{Phi11}), one can readily get
\begin{equation}
\label{eq-F}
1+\sum_{k=1}^{N}\frac{(\lambda_{j}^{2}\lambda_{k}^{\ast}+\lambda_{j}\lambda_{k}^{\ast
2}c_{j}c_{k}^{\ast})}{(\lambda^{2}_{j}-\lambda^{\ast
2}_{k})}F_{k}=0,
\end{equation}
where
\begin{equation}
\label{F}
F_{k}=\frac{2\psi_{2}^{\ast}(\lambda_{k})}{\lambda_{k}^{\ast
2}\dot{a}^{\ast}(\lambda_{k})}\mathrm{e}^{i\lambda_{k}^{\ast 2}x},
\end{equation}
and the time dependence of $b_{j}$ in (\ref{dyn2}) is explicitly
taken into account, so that
\begin{equation}
\label{c-original} c_{j}=b_{j}\exp
\left(2i\lambda_{j}^{2}x-i/(2\lambda^{2}_{j})t \right).
\end{equation}
Using the expression for the $N$-soliton solution (\ref{u_sol})
and taking into account (\ref{F}) we have for $u$
\begin{equation}
\label{u-F} u=\sum_{k=1}^{N}c_{k}^{\ast}F_{k}.
\end{equation}
 From (\ref{eq-F}) and (\ref{u-F}) one can obtain the $N$-soliton solution $u$ in a
compact form
\begin{equation}
\label{N-soliton}
u=\sum_{k,j=1}^{N}c_{k}^{\ast}(\mathbf{K}^{-1})_{kj},
\end{equation}
where the elements of the $N\times N$ matrix $\mathbf{K}$ are
\begin{equation}
\label{K-matrix}
K_{jk}=\frac{\lambda_{j}\lambda_{k}^{\ast}}{\lambda_{k}^{\ast
2}-\lambda_{j}^{ 2}}(\lambda_{j}+ \lambda_{k}^{\ast}c_{j}
c_{k}^{\ast}).
\end{equation}
Equations (\ref{N-soliton}) and (\ref{K-matrix}) were previously
obtained by Lenells \cite{Lenells_N-soliton2010} using the
dressing method, but the solution $u$ was not expressed in the
determinant form. Note also that in \cite{Lenells_N-soliton2010},
a more general equation than (\ref{main}) was considered. Using
(\ref{N-soliton}) and the identity
\begin{equation}
\mathbf{A}_{1}^{\mathrm{T}}\mathbf{A}^{-1}\mathbf{A}_{2}=\frac{\det
(\mathbf{A}+\mathbf{A}_{2}\mathbf{A}_{1}^{\mathrm{T}})}{\det
(\mathbf{A})}-1,
\end{equation}
where $\mathbf{A}$ is an arbitrary $N\times N$ matrix,
$\mathbf{A}_{1}$ and $\mathbf{A}_{2}$ are arbitrary $N\times 1$
matrices respectively, we can write the $N$-soliton solution of
equation (\ref{main}) as
\begin{equation}
\label{u-det} u=\frac{\det (\tilde{\mathbf{K}})-\det
(\mathbf{K})}{\det (\mathbf{K})},
\end{equation}
where the elements of the matrix $\tilde{\mathbf{K}}$ are
\begin{equation}
\label{K_tilde} \tilde{K}_{jk}=K_{jk}+c^{\ast}_{k}.
\end{equation}
In what follows, we parameterize the complex numbers $\lambda_{j}$
and $b_{j}$ in terms of four real parameters $\Delta_{j}>0$,
$0<\gamma_{j}<\pi$, $x_{0 j}$ (the initial position of the
soliton) and $\phi_{0 j}$ (the initial phase) as
\begin{eqnarray}
\label{parametrize1}\lambda_{j}^{2}=\Delta^{2}_{j}(\cos\gamma_{j}
+i\sin\gamma_{j}),
\\
\label{parametrize2} b_{j}=\exp (2x_{0
j}\Delta^{2}_{j}\sin\gamma_{j}+i\phi_{0 j}).
\end{eqnarray}
With this parametrization $\lambda_{j}$ and $-\lambda_{j}$ lie in
the 1-st and 3-rd quadrants respectively of the complex plane
($\pm \lambda_{j}^{\ast}$ -- in the 2-st and 4-th quadrants
respectively). Then $c_{j}$ determined by (\ref{c-original}) can
be written as
\begin{equation}
\label{c_j} c_{j}=\exp (-z_{j}+i\Phi_{j}),
\end{equation}
where
\begin{equation}
\label{z} z_{j}=2\Delta^{2}_{j}\left (x-x_{0
j}+v_{j}t\right)\sin\gamma_{j}  ,
\end{equation}
with $v_{j}=1/(4\Delta^{4}_{j})$, and
\begin{equation} \label{Fi} \Phi_{j}=2\Delta^{2}_{j}
\left(x-v_{j}t\right)\cos\gamma_{j} +\phi_{0 j}.
\end{equation}
Using this parametrization and (\ref{K-matrix}), for the elements
$K_{jk}$ of the matrix $\mathbf{K}$ one can obtain
\begin{equation}
\label{K_jk}
K_{jk}=\frac{\Delta_{j}\Delta_{k}\left[\Delta_{j}\mathrm{e}^{i(\gamma_{j}-\gamma_{k}/2)}
+\Delta_{k}\mathrm{e}^{i(\gamma_{j}/2-\gamma_{k})}
\mathrm{e}^{-z_{j}-z_{k}+i(\Phi_{j}-\Phi_{k})}\right]}{\Delta_{k}^{2}
\mathrm{e}^{-i\gamma_{k}}-\Delta_{j}^{2}\mathrm{e}^{i\gamma_{j}}}.
\end{equation}
In particular, for $K_{jj}$ we have
\begin{equation}
\label{K_jj} K_{jj}=\frac{i\Delta_{j}\mathrm{e}^{-z_{j}}\cosh
(z_{j}+i\gamma_{j}/2)}{\sin\gamma_{j}}.
\end{equation}

\subsection{N-soliton solutions: rational and a
mixture of exponential and rational solutions}

Equation (\ref{u-det}) with matrix elements $K_{jk}$ and $c_{j}$
determined by (\ref{K_jk}) and (\ref{c_j}) respectively is the
$N$-soliton solution in exponential functions that decays
exponentially at infinity. It is essential that for these
solutions $\gamma_{j}<\pi$ and $\gamma_{k}<\pi$ in (\ref{K_jk}).
However, the apparent singularity $\gamma_{j}=\pi$ for $K_{jj}$ in
(\ref{K_jj}) is fictitious and it is easy to show that in the
limit $\gamma_{j}\rightarrow\pi$ the elements $K_{jj}$ become
rational functions of $x$. In the limit $\gamma_{j}\rightarrow\pi$
and $\gamma_{k}\rightarrow\pi$, for the values $c_{j}^{\ast}$ in
(\ref{c_j}) and $K_{jj}$ in (\ref{K_jj}) one can obtain
\begin{equation}
\label{c_algeb}
c_{j}^{\ast}=\exp\left[2i\Delta_{j}^{2}\left(x-v_{j}t\right)-i\phi_{0
j}\right],
\end{equation}
and
\begin{equation}
\label{K_jj_algeb}
K_{jj}=\frac{1}{2}\left[i\Delta_{j}-4\Delta_{j}^{3}\left(x-x_{0
j}+v_{j}t\right)\right],
\end{equation}
and for $j\neq k$
\begin{equation}
\label{K_jk_algeb} K_{jk}=\frac{i\Delta_{j}\Delta_{k}}
{\Delta_{k}^{2}-\Delta_{j}^{2}}\left\{\Delta_{k}
\exp\left[2i\left(\Delta_{k}^{2}-\Delta_{j}^{2}\right)\left(x+\frac{t}
{4\Delta_{j}^{2}\Delta_{k}^{2}}\right)\right] -\Delta_{j}\right\}.
\end{equation}
Equation (\ref{u-det}) with $c_{j}$ and the matrix elements
$K_{jk}$ determined by (\ref{c_algeb}) and (\ref{K_jj_algeb}),
(\ref{K_jk_algeb}) respectively is the rational $N$-soliton
solution decaying in power law at infinity.

An interesting situation arises if $\gamma_{j}\rightarrow\pi$ with
$j=1\dots M$, where $M<N$ and $\gamma_{k}\neq\pi$ with $k=M+1\dots
N$. Then from (\ref{K_jk}) one can find
\begin{equation}
\label{K_jk_mix}
K_{jk}=\frac{\Delta_{j}\Delta_{k}}{\Delta_{k}^{2}+\Delta_{j}^{2}\mathrm{e}^{i\gamma_{k}}}
\left(i\Delta_{k}\mathrm{e}^{-z_{k}+i\Psi_{jk}}-\Delta_{j}\mathrm{e}^{i\gamma_{k}/2}\right),
\end{equation}
where
\begin{equation}
\Psi_{jk}=-2(\Delta_{j}^{2}+\Delta_{k}^{2}\cos\gamma_{k})x+\left(\frac{1}{\Delta_{j}^{2}}
+\frac{\cos\gamma_{k}}{\Delta_{k}^{2}}\right)\frac{t}{2}.
\end{equation}
Similarly, if $\gamma_{k}\rightarrow\pi$ and $\gamma_{j}\neq\pi$
(note that the matrix $K_{jk}$ is not symmetric) we have
\begin{equation}
K_{jk}=\frac{\Delta_{j}\Delta_{k}}{\Delta_{k}^{2}\mathrm{e}^{-i\gamma_{j}}+\Delta_{j}^{2}}
\left(i\Delta_{j}+\Delta_{k}\mathrm{e}^{-i\gamma_{j}/2}\mathrm{e}^{-z_{j}+i\Psi_{jk}}\right),
\end{equation}
where
\begin{equation}
\Psi_{jk}=2(\Delta_{k}^{2}+\Delta_{j}^{2}\cos\gamma_{j})x-\left(\frac{1}{\Delta_{k}^{2}}
+\frac{\cos\gamma_{j}}{\Delta_{j}^{2}}\right)\frac{t}{2}.
\end{equation}
Equation (\ref{u-det}) with the coefficients $c_{j}$ and $c_{k}$
determined by (\ref{c-original}) and (\ref{c_algeb}) respectively,
and the matrix elements determined by (\ref{K_jk}) and
(\ref{K_jk_mix}), is an $N$-soliton solution consisting of a
mixture of $M$ rational and $N-M$ exponential functions.

\subsection{One-soliton solutions: exponential and rational}

The case $N=1$ corresponds to the one-soliton solution of equation
(\ref{main}). Then from (\ref{K-matrix}) and (\ref{K_tilde}) we
have
\begin{equation}
\label{K11-two}
K_{11}=\frac{|\lambda_{1}|^{2}(\lambda_{1}+\lambda^{\ast}_{1}|c_{1}|^{2})}{\lambda_{1}^{\ast
2}-\lambda_{1}^{2}}, \quad \tilde{K}_{11}=K_{11}+c^{\ast}_{1},
\end{equation}
and from (\ref{u-det}) one can readily get
\begin{equation}
\label{u-1sol}
u_{1}=\frac{c_{1}^{\ast}}{K_{11}}=\frac{c_{1}^{\ast}(\lambda_{1}^{\ast
2}-\lambda_{1}^{2})}{|\lambda|^{2}(\lambda_{1}+\lambda^{\ast}_{1}|c_{1}|^{2})}.
\end{equation}
Using the parametrization (\ref{parametrize1}) and
(\ref{parametrize2}) this solution takes the form
\begin{equation}
\label{one-soliton1} u_{1}=\frac{\sin\gamma_{1}\exp
(-i\Phi_{1})}{i\Delta_{1}\cosh (z_{1}+i\gamma_{1}/2)}.
\end{equation}
An explicit expression for $u_{1}$ in terms of the soliton
amplitude and phase is
\begin{equation}
\label{one-soliton2} u_{1}=\frac{\sin\gamma_{1}\exp
\{-i\Phi_{1}-i\arctan[\tanh z_{1}\tan
(\gamma_{1}/2)]\}}{i\Delta_{1}\sqrt{\cosh^{2}z_{1}-\sin^{2}(\gamma_{1}/2)}}.
\end{equation}
Earlier this solution was obtained by Davydova and Lashkin
\cite{Lashkin1991,Lashkin1994} without using the IST. The soliton
velocity (in the negative direction of $x$-axis) $v_{1}$,
amplitude $A$ and the characteristic halfwidth of the soliton $w$
are
\begin{equation}
\label{velocity} v_{1}=\frac{1}{4\Delta_{1}^{4}},\quad
A=\frac{\sin\gamma_{1}}{\Delta_{1}}, \quad
w=\frac{1}{2\Delta_{1}^{2}\sin\gamma_{1}}.
\end{equation}
It is seen that the soliton can not be motionless, and it moves
only in the negative direction of $x$-axis. In the limit
$\gamma_{1}\rightarrow \pi$, from (\ref{one-soliton1}) (or,
directly from (\ref{c_algeb}), (\ref{K_jj_algeb}) and
(\ref{u-1sol}) ) one can obtain the soliton with algebraic decay
at infinity,
\begin{equation}
\label{algebraic} u_{1}=\frac{2\exp
(-i\Phi_{1})}{\Delta_{1}(i-2y)},
\end{equation}
where $y=2\Delta_{1}(x-x_{0}+v_{1}t)$ and
$\Phi_{1}=-2\Delta_{1}^{2}(x-v_{1}t)+\phi_{01}$. In terms of the
amplitude and phase,  the expression (\ref{algebraic}) takes the
form
\begin{equation}
\label{algebraic1} u_{1}=\frac{2\exp
[-i\Phi_{1}+i\,\mathrm{arccot}\,(4\Delta_{1}^{2}y)]}
{\Delta_{1}\sqrt{1+16\Delta_{1}^{4}y^{2}}}.
\end{equation}
This algebraic soliton solution of the DLFL equation (\ref{main})
was first obtained in \cite{Lashkin1994} and then rediscovered in
\cite{Lenells2009_Nonlinearity}.

\subsection{Two-soliton solutions}

In the case $N=2$ the corresponding matrix elements in
(\ref{K-matrix}) and (\ref{K_tilde}) have the form
\begin{eqnarray}
K_{12}=\frac{\lambda_{1}\lambda_{2}^{\ast}(\lambda_{1}+\lambda_{2}^{\ast}
c_{1}c_{2}^{\ast})}{\lambda_{2}^{\ast 2}-\lambda_{1}^{2}}, \quad
\tilde{K}_{12}=K_{12}+c^{\ast}_{2}, \label{K12_2}
\\
K_{21}=\frac{\lambda_{2}\lambda_{1}^{\ast}(\lambda_{2}+\lambda_{1}^{\ast}
c_{2}c_{1}^{\ast})}{\lambda_{1}^{\ast 2}-\lambda_{2}^{2}}, \quad
\tilde{K}_{21}=K_{21}+c^{\ast}_{1},
\label{K21_2} \\
K_{22}=\frac{|\lambda_{2}|^{2}(\lambda_{2}+\lambda^{\ast}_{2}|c_{2}|^{2})}{\lambda_{2}^{\ast
2}-\lambda_{2}^{2}}, \quad \tilde{K}_{22}=K_{22}+c^{\ast}_{2},
\label{K22_2}
\end{eqnarray}
and $K_{11}$ is determined by (\ref{K11-two}). Then, as one can
see from  (\ref{u-det}), the corresponding general two-soliton
solution is
\begin{equation}
\label{u-2sol}
u_{2}=\frac{c_{1}^{\ast}(K_{22}-K_{12})+c_{2}^{\ast}(K_{11}-K_{21})}{K_{11}K_{22}-K_{12}K_{21}},
\end{equation}
where $c_{1}^{\ast}$ and $c_{2}^{\ast}$  are determined by
(\ref{c_j}).

In the particular case, when the eigenvalues $\lambda_{1,2}^{2}$
are purely imaginary (this corresponds to $\gamma_{1,2}=\pi/2$),
we have
\begin{equation}
c_{1}=\mathrm{e}^{-y_{1}}, \quad c_{2}=\mathrm{e}^{-y_{2}},
\end{equation}
\begin{equation}
K_{11}=i\Delta_{1}\mathrm{e}^{-y_{1}}\cosh (y_{1}+i\pi/4), \quad
 K_{22}=i\Delta_{2}\mathrm{e}^{-y_{2}}\cosh (y_{2}+i\pi/4),
\end{equation}
\begin{equation}
K_{12}=
i\frac{\Delta_{1}\Delta_{2}}{\Delta_{1}^{2}+\Delta_{2}^{2}}\left(\Delta_{1}\mathrm{e}^{i\pi/4}+
\Delta_{2}\mathrm{e}^{-i\pi/4}\mathrm{e}^{-y_{1}-y_{2}}\right),
\end{equation}
\begin{equation}
K_{21}=i\frac{\Delta_{1}\Delta_{2}}{\Delta_{1}^{2}+\Delta_{2}^{2}}\left(\Delta_{2}\mathrm{e}^{i\pi/4}+
\Delta_{1}\mathrm{e}^{-i\pi/4}\mathrm{e}^{-y_{1}-y_{2}}\right),
\end{equation}
where  $j=1,2$ and $y_{j}=2\Delta_{j}^{2}(x-x_{0j}+v_{j}t)$, and
the corresponding two-soliton solution (\ref{u-2sol}) has the
simple form
\begin{equation}
u=\frac{i(\Delta_{2}^{2}-\Delta_{1}^{2})[\Delta_{1}\cosh
y_{1}^{+}-\Delta_{2}\cosh
y_{2}^{+}]}{(\Delta_{1}^{2}+\Delta_{2}^{2})\{\Delta_{1}\Delta_{2}
[\cosh y_{1}^{+}\cosh y_{2}^{+}+2i\sinh(y_{1}+y_{2})]
+\Delta_{1}^{2}+\Delta_{2}^{2}\}},
\end{equation}
where $y^{+}_{j}=y_{j}+i\pi/4$.

As another particular example, consider the two-soliton
rational-exponential bound state. If the velocities $v_{1}$ and
$v_{2}$ of the components in a two-soliton solution are equal,
then the solution represents a bound state. Consider such a
solution when one of the components is an algebraic soliton. Let
$\gamma_{1}\rightarrow\pi$ and $\gamma_{2}\equiv\gamma <\pi$, and
$v_{1}=v_{2}\equiv v$, $x_{0 1}=x_{0 2}=0$, $\phi_{0 1}=\phi_{0
2}=0$. Then the corresponding coefficients $c_{1}^{\ast}$ and
$c_{2}^{\ast}$ are
\begin{equation}
c_{1}^{\ast}=\exp[2i\Delta^{2}(x-vt)], \quad
c_{2}^{\ast}=\exp[-y\sin\gamma-2i\Delta^{2}(x-vt)\cos\gamma],
\end{equation}
where $y=2\Delta^{2}(x+vt)$. From (\ref{K_jj_algeb}) and
(\ref{K_jj}) we have
\begin{equation}
K_{11}=\frac{\Delta}{2}(i-2y), \quad
 K_{22}=\frac{i\Delta\mathrm{e}^{-y\sin\gamma}\cosh
(y\sin\gamma+i\gamma/2)}{\sin\gamma},
\end{equation}
and from (\ref{K_jk_mix}) one can obtain
\begin{equation}
K_{12}=\frac{\Delta
\left(i\mathrm{e}^{-y\sin\gamma+i\Psi-i\gamma/2}-1\right)}{2\cos
(\gamma/2)} , \quad  K_{21}=\frac{\Delta
\left(i\mathrm{e}^{i\gamma/2}+\mathrm{e}^{-y\sin\gamma-i\Psi}\right)}{2\cos
(\gamma/2)},
\end{equation}
where $\Psi=qx-\Omega t$ with
$\Omega=\cos^{2}(\gamma/2)/\Delta^{2}$ and $q=\Omega/v$. The
solution (\ref{u-2sol})  with such coefficients and matrix
elements pulsates with two independent frequencies $\Omega$ and
$1/(2\Delta^{2})$. Note that, despite the presence of an
exponential component, the solution decays at infinity by a power
law $\sim 1/|y|$. In particular, for $\gamma=\pi/2$ (this
corresponds to the largest amplitude of the exponential
component), the solution simplified to
\begin{equation}
u=\frac{i\mathrm{e}^{i\Psi}\cosh
y^{+}-y+(3/2)i}{\Delta[2i(i-2y)\cosh y^{+}-\cosh y^{-}+\cos\Psi]},
\end{equation}
where $y^{\pm}=y\pm i\pi/4$. This solution is a breather (there is
one independent frequency) and oscillates with a period $T=4\pi
\Delta^{2}$.

\section{ \label {Sec4} The asymptotic behavior of the $N$-soliton solution}
Consider the time asymptotics of the two-soliton solution
(\ref{u-2sol}), assuming that the velocities of the two soliton
components $v_{1}$ and $v_{2}$ are different. Assume $v_{1}>v_{2}$
and let $z_{1}$ be fixed. Then at $t\rightarrow -\infty$ we have
$z_{2}\rightarrow\infty$ and $|c_{1}|$ is finite while
$|c_{2}|\rightarrow 0$. Evaluating the corresponding $K_{jk}$ from
(\ref{K11-two}) and (\ref{K12_2})-(\ref{K22_2}) and inserting into
 (\ref{u-2sol}) one can obtain the leading term as
\begin{equation}
\label {u-lead1} u_{2}\sim \frac{c_{1}^{\ast}(\lambda_{2}^{\ast
2}-\lambda_{1})(\lambda_{1}^{\ast
2}-\lambda_{2})[|\lambda_{2}|^{2}\lambda_{2}(\lambda_{2}^{\ast
2}-\lambda_{1}^{2})-\lambda_{1}^{2}\lambda_{2}^{\ast}(\lambda_{2}^{\ast
2}-\lambda_{2}^{2})]}{\lambda_{2}|\lambda_{1}|^{2}|\lambda_{2}|^{2}
[(\lambda_{2}^{\ast 2}-\lambda_{1})(\lambda_{1}^{\ast
2}-\lambda_{2})
(\lambda_{1}+\lambda_{1}|c_{1}|^{2})-\lambda_{1}(\lambda_{1}^{\ast
2}-\lambda_{1})(\lambda_{2}^{\ast 2}-\lambda_{2})]} ,
\end{equation}
that can be written in the form
\begin{equation}
\label{u-mimic}
u_{2}\sim\frac{\tilde{c}_{1}^{\ast}(\lambda_{1}^{\ast
2}-\lambda_{1}^{2})}{|\lambda|^{2}(\lambda_{1}+\lambda^{\ast}_{1}|\tilde{c}_{1}|^{2})},
\end{equation}
where
\begin{equation}
\tilde{c}_{1}=c_{1}\exp\left[-\ln\frac{\lambda_{2}^{2}
(\lambda_{2}^{\ast 2}-\lambda_{1}^{2})}{\lambda_{2}^{\ast 2}
(\lambda_{2}^{2}-\lambda_{1}^{2})}\right].
\end{equation}
One can see that the asymptotic of $u_{2}$ determined by
(\ref{u-mimic}) has the same form as the one-soliton solution
(\ref{u-1sol}) except the phase shifts, so that we have
\begin{equation}
u_{2}\sim u_{1}(z_{1}+\Delta z_{1}^{-},\Phi_{1}+\Delta
\Phi_{1}^{-}) ,
\end{equation}
where
\begin{equation}
\label{z-minus} \Delta z_{1}^{-}=\ln\left |\frac{\lambda_{2}^{\ast
2}-\lambda_{1}^{2}}{\lambda_{2}^{2}-\lambda_{1}^{2}}\right| ,\quad
\Delta \Phi_{1}^{-}=-\arg\frac{\lambda_{2}^{\ast
2}-\lambda_{1}^{2}}{\lambda_{2}^{2}-\lambda_{1}^{2}}-\arg
\frac{\lambda_{2}^{2}}{\lambda_{2}^{\ast 2}}+\pi .
\end{equation}
Similarly, if $t\rightarrow +\infty$ we have $z_{2}\rightarrow
-\infty$ and then $|c_{1}|$ is finite while $|c_{2}|\rightarrow
\infty$. The leading term in that case is
\begin{equation}
u_{2}\sim\frac{c_{1}^{\ast}\lambda_{2}^{\ast 2}(\lambda_{1}^{\ast
2}-\lambda_{1}^{2})(\lambda_{2}^{2}-\lambda_{1}^{2})(\lambda_{1}^{\ast
2}-\lambda_{2}^{2})}{\lambda_{2}^{2}|\lambda_{1}|^{2}
[\lambda_{1}(\lambda_{1}^{2}-\lambda_{2}^{\ast
2})(\lambda_{1}^{\ast 2}-\lambda_{2}^{\ast 2})
+\lambda_{1}^{\ast}(\lambda_{1}^{2}-\lambda_{2}^{\ast
2})(\lambda_{1}^{\ast 2}-\lambda_{1}^{2})|c_{1}|^{2}]} ,
\end{equation}
and it can be written as (\ref{u-mimic}), where
\begin{equation}
\tilde{c}_{1}=c_{1}\exp\left[\ln\frac{\lambda_{2}^{2}
(\lambda_{2}^{\ast 2}-\lambda_{1}^{2})}{\lambda_{2}^{\ast 2}
(\lambda_{2}^{2}-\lambda_{1}^{2})}\right].
\end{equation}
And
\begin{equation}
u_{2}\sim u_{1}(z_{1}+\Delta z_{1}^{+},\Phi_{1}+\Delta
\Phi_{1}^{+}) ,
\end{equation}
where
\begin{equation}
\label{z-plus} \Delta z_{1}^{+}=-\ln\left |\frac{\lambda_{2}^{\ast
2}-\lambda_{1}^{2}}{\lambda_{2}^{2}-\lambda_{1}^{2}}\right| ,\quad
\Delta \Phi_{1}^{+}=\arg\frac{\lambda_{2}^{\ast
2}-\lambda_{1}^{2}}{\lambda_{2}^{2}-\lambda_{1}^{2}}+\arg
\frac{\lambda_{2}^{2}}{\lambda_{2}^{\ast 2}}+\pi .
\end{equation}
The total shifts are determined by $\Delta z_{1}=\Delta
z_{1}^{+}-\Delta z_{1}^{-}$ and $\Delta \Phi_{1}=\Delta
\Phi_{1}^{+}-\Delta \Phi_{1}^{-}$. Then, taking into account
(\ref{z}), (\ref{z-minus}) and (\ref{z-plus}), we have for the
position shift of the soliton $j=1$,
\begin{equation}
\Delta x_{1}=\frac{2i}{\lambda_{1}^{ 2}-\lambda_{1}^{\ast
2}}\ln\left |\frac{\lambda_{2}^{\ast
2}-\lambda_{1}^{2}}{\lambda_{2}^{2}-\lambda_{1}^{2}}\right|
\end{equation}
and for the corresponding phase shift,
\begin{equation}
\Delta \Phi_{1}=2\arg\frac{\lambda_{2}^{\ast
2}-\lambda_{1}^{2}}{\lambda_{2}^{2}-\lambda_{1}^{2}}+2\arg
\frac{\lambda_{2}^{2}}{\lambda_{2}^{\ast 2}}.
\end{equation}
If we now fix $z_{2}$ (as before $v_{1}>v_{2}$), then after
similar calculations one can obtain
\begin{equation}
\Delta x_{2}=-\frac{2i}{\lambda_{2}^{ 2}-\lambda_{2}^{\ast
2}}\ln\left |\frac{\lambda_{1}^{\ast
2}-\lambda_{2}^{2}}{\lambda_{1}^{2}-\lambda_{2}^{2}}\right|,
\end{equation}
and for the corresponding phase shift,
\begin{equation}
\Delta \Phi_{2}=-2\arg\frac{\lambda_{1}^{\ast
2}-\lambda_{2}^{2}}{\lambda_{1}^{2}-\lambda_{2}^{2}}-2\arg
\frac{\lambda_{1}^{2}}{\lambda_{1}^{\ast 2}}.
\end{equation}
Generalization to the $N$-soliton solution can be performed
straightforwardly following the two-soliton case. Soliton
velocities are assumed to be ordered as $v_{1}>v_{2}>\dots>v_{N}$.
As in the two-soliton case, we consider the limits $t\rightarrow
-\infty$ and $t\rightarrow \infty$ and assume that $z_{n}$ is
fixed. Then in the first case we have $|c_{j}|\rightarrow\infty$
for $j<n$ and $|c_{j}|\rightarrow  0$ for $n<j$, and in the second
$|c_{j}|\rightarrow 0$ for $j<n$ and $|c_{j}|\rightarrow \infty$
for $n<j$. In both cases, the leading-order asymptotics of the
matrices $\tilde{\mathbf{K}}$ and $\mathbf{K}$ are Cauchy matrices
\cite{Vein} and the corresponding determinants in (\ref{u-det})
are factorized (that is, they are Cauchy determinants). Omitting
the calculations, we present only the final results. The
asymptotic form of the $N$-soliton solution is
\begin{equation}
u_{N}\sim u_{1}(z_{n}+\Delta z_{n}^{-},\Phi_{n}+\Delta
\Phi_{n}^{-}),
\end{equation}
as $t\rightarrow -\infty$, and
\begin{equation}
u_{N}\sim u_{1}(z_{n}+\Delta z_{n}^{+},\Phi_{n}+\Delta
\Phi_{n}^{+}),
\end{equation}
as $t\rightarrow \infty$ with $\Delta z_{n}^{+} =-\Delta
z_{n}^{-}$ and $\Delta \Phi_{n}^{+} =-\Delta\Phi_{n}^{-}+2\pi N$.
The total phase shifts of the $n$-th soliton
\begin{equation}
\Delta x_{n}=\frac{2i}{\lambda_{n}^{ 2}-\lambda_{n}^{\ast 2}}
\left[\sum_{j=n+1}^{N}\ln\left |\frac{\lambda_{j}^{\ast
2}-\lambda_{n}^{2}}{\lambda_{j}^{2}-\lambda_{n}^{2}}\right|-\sum_{j=1}^{n-1}\ln\left
|\frac{\lambda_{j}^{\ast
2}-\lambda_{n}^{2}}{\lambda_{j}^{2}-\lambda_{n}^{2}}\right|\right],
\end{equation}

\begin{eqnarray}
\Delta \Phi_{n}=2 \sum_{j=n+1}^{N}\left[\arg\left
(\frac{\lambda_{j}^{\ast
2}-\lambda_{n}^{2}}{\lambda_{j}^{2}-\lambda_{n}^{2}}\right)+\arg\left(\frac{\lambda_{j}^{2}}{\lambda_{j}^{\ast
2}}\right)\right] \nonumber  \\
 -2\sum_{j=1}^{n-1}\left[\arg\left
(\frac{\lambda_{j}^{\ast
2}-\lambda_{n}^{2}}{\lambda_{j}^{2}-\lambda_{n}^{2}}\right)+\arg\left(\frac{\lambda_{j}^{2}}{\lambda_{j}^{\ast
2}}\right)\right].
\end{eqnarray}
In the general case, the asymptotic $N$-soliton solution is a
superposition of $N$ separate one-soliton solutions with the
corresponding parameters $\Delta_{j}$ and $\gamma_{j}$ where
$j=1\dots N$. Note that, for a rational soliton
($\gamma_{n}\rightarrow\pi$)  we have
$\lambda_{n}^{2}=-\Delta_{n}^{2}$ and, using L'H\^{o}pital's rule,
we can obtain $\Delta x_{n}=0$, so that the position shift of the
of this soliton upon interaction with other solitons is equal to
zero.

\section{\label {Sec5} Conclusion}

In this paper, we have presented a simple and constructive method
for finding $N$-soliton solutions of the DLFL equation
(\ref{main}) to describe the dynamics of nonlinear ion-cyclotron
waves in a plasma. The proposed method is based on the classical
formulation of the IST and differs from the Hirota bilinear method
used in \cite{Matsuno_bright2012} as well as the dressing method
in \cite{Lenells_N-soliton2010} primarily in that it allows one to
take into account the contribution of the continuous spectrum that
is, the radiation field. The resulting general expression for
arbitrary initial data decaying at infinity is written in terms of
discrete and continuous scattering data and the corresponding Jost
solutions and consists of soliton and nonsoliton (radiative)
parts. The first of them corresponds to the discrete spectrum of
the spectral problem (\ref{spec1}) and the second part does to the
continuous spectrum. The radiation part is represented as an
integral over the spectral parameter, and depends on one of the
Jost solutions and the reflection coefficient. Thus, the radiative
part corresponding to quasilinear ion-cyclotron waves can, in
principle, be determined explicitly if the corresponding Jost
solution and the reflection coefficient are known. For example,
under certain conditions, that is, using perturbation theory, the
Jost solution and coefficient $a(\lambda)$ can be taken as purely
soliton ones.

We have found two new types of $N$-soliton solutions the DLFL
equation (\ref{main}): an algebraic $N$-soliton solution in
rational functions, and a solution in the form of a mixture of $M$
rational and $N-M$ exponential functions. Both solutions are
presented in determinant form. As an example, we write out two
two-soliton solutions explicitly. The first of them corresponds to
purely imaginary eigenvalues, and the second represents a solution
in the form of a bound state of the usual bright soliton and the
algebraic soliton, which pulsates with two independent
frequencies.

\section{Declaration of competing interest}
The authors declare that they have no known competing financial
interests or personal relationships that could have appeared to
influence the work reported in this paper.

\section{CRediT authorship contribution statement}
\textbf{V. M. Lashkin}: Conceptualization, Methodology,
Validation, Formal analysis, Investigation.

\section{Acknowledgments}

The work was supported by the National Research Foundation of
Ukraine, grant 2020.02/0015.

\section{Appendix}

In this appendix we give a short outline of the derivation of the
two-dimensional nonlinear equation describing the dynamics of
ion-cyclotron waves in plasmas, first suggested by Davydova and
Lashkin in \cite{Lashkin1991} (see also \cite{Lashkin1994}), which
in the one-dimensional case reduces to the DLFL equation
(\ref{main}). For a plasma in an uniform external magnetic field
$\mathbf{H}_{0}=H_{0}\hat{\mathbf{z}}$ oriented along the
$z$-axis, the linear dispersion relation for the electrostatic
ion-cyclotron waves (the Bernstein modes) in the short-wavelength
limit $k_{\perp}\rho_{i}\gg 1$ under the conditions
$k_{\perp}\rho_{e}\ll 1$ and $\omega\ll k_{z}v_{Te}$ is,
\begin{equation}
\label{disp_phys} \omega
(\mathbf{k})=n\Omega_{i}\left[1+\frac{1}{\sqrt{2\pi}(1
+T_{i}/T_{e})k_{\perp}\rho_{i}}\right] \equiv
n\Omega_{i}[1+R(k_{\perp})],
\end{equation}
where $R(k_{\perp})\ll 1$ \cite{Akhiezer_book1975}. Here $\omega$
and $\mathbf{k}$ are the frequency and wave vector
respectively,$k_{\perp}=\sqrt{k_{x}^{2}+k_{y}^{2}}$, $\Omega_{i}$
is the ion-cyclotron frequency, $\rho_{\alpha}$, $v_{T\alpha}$ and
$T_{\alpha}$ are the Larmor radius, thermal velocity and
temperature of particle species $\alpha$ ($e$ for electrons and
$i$ for ions) respectively, $n=1,2,\dots$ . Next, we consider the
case of only the lowest harmonic $n=1$. The Maxwell equation
$\nabla\cdot\mathbf{D}=0$ for the electrical displacement
$\mathbf{D}(\omega,\mathbf{k})=\varepsilon(\omega,\mathbf{k})\mathbf{E}(\omega,\mathbf{k})$,
where $\varepsilon$ and $\mathbf{E}$ are the dielectric function
and electric field in the Fourier space respectively, can be
written in the physical two-dimensional space as
\begin{equation}
\label{Maxvell} \nabla_{\perp}\cdot (\hat{\varepsilon}
\nabla_{\perp}\varphi)=0,
\end{equation}
where $\hat{\varepsilon}$ is considered as a differential operator
with $\omega\rightarrow i\partial/\partial t$ and
$\mathbf{k}\rightarrow -i\nabla_{\perp}\equiv -i(\partial/\partial
x,\partial/\partial y)$. The principal nonlinear effect for
ion-cyclotron waves is the perturbation of the magnetic field
$\delta H_{z}$ \cite{Lashkin1991,Lashkin1994}. In this case, the
nonlinear correction to the ion-cyclotron frequency in the
expression for $\hat{\varepsilon}$ is taken into account, so that
$\Omega_{i}\rightarrow \Omega_{i}(1+h)$, where the relative
nonlinear perturbation $h$ of the magnetic field $H_{0}$ is
\begin{equation}
\label{B} h=\frac{\delta
H_{z}}{H_{0}}=-\frac{\omega_{pe}^{2}m|\Phi|^{2}}{4H_{0}^{2}T_{e}},
\end{equation}
where $\Phi$ is the envelope of the electrostatic potential
$\tilde{\Phi}$ at the ion-cyclotron frequency,
\begin{equation}
\label{envelope} \tilde{\Phi}=\frac{1}{2}[\Phi\exp
(-i\Omega_{i}t)+\mathrm{c}.\,\mathrm{c}],
\end{equation}
and $\omega_{pe}$ is the electron plasma frequency, $m$ is the
electron mass. In \cite{Lashkin1991,Lashkin1994}, the anisotropy
of electron temperatures was also taken into account, and then for
the electron temperature $T_{e}$ in (\ref{B}) it would be
$T_{e}\rightarrow T_{e,\parallel}^{2}/T_{e,\perp}$, where
$T_{e,\parallel}$ (i.e., along the $z$-axis) and $T_{e,\perp}$ are
parallel and transverse electron temperatures respectively.
Expanding $\varepsilon (\omega,\mathbf{k})$  near the
eigenfrequency $\omega_{\mathbf{k}}$ determined by
(\ref{disp_phys}) with the nonlinear correction (\ref{B}) yields
\begin{equation}
\label{var-eps} \varepsilon (\omega,\mathbf{k})=\varepsilon
(\omega_{\mathbf{k}},\mathbf{k})+
\varepsilon^{\prime}(\omega_{\mathbf{k}},\mathbf{k})(\omega-\omega_{\mathbf{k}}),
\end{equation}
where $\varepsilon^{\prime}(\omega_{k})\equiv\partial \varepsilon
(\omega)/\partial\omega\mid_{\omega=\omega_{k}}$. Substituting
(\ref{var-eps}) into (\ref{Maxvell}) along with (\ref{disp_phys})
and (\ref{B}), one can obtain the nonlinear equation
\cite{Lashkin1991,Lashkin1994} in the form
\begin{equation}
\label{equation2D}
\Delta_{\perp}\left(\frac{i}{\Omega_{i}}\frac{\partial\Phi}{\partial
t}-\hat{R}\Phi\right)=\nabla_{\perp}\cdot (h\nabla\Phi),
\end{equation}
where $\Delta_{\perp}=\partial^{2}/\partial
x^{2}+\partial^{2}/\partial y^{2}$ and the operator $\hat{R}$ is
defined by
\begin{equation}
\label{R} \hat{R}\Phi (\mathbf{r},t)=\int
R(k_{\perp})\hat{\Phi}(\mathbf{k}_{\perp},t)\exp
(i\mathbf{k}_{\perp}\cdot \mathbf{r})\,d\mathbf{k}_{\perp}.
\end{equation}
In the one-dimensional case, and in the dimensionless variables
\begin{equation}
x\rightarrow \frac{x}{\sqrt{2\pi}(1+T_{i}/T_{e})\rho_{i}},\quad
u\rightarrow \Phi\frac{\omega_{pe}}{2H_{0}}\sqrt{\frac{m}{T_{e}}}
,
\end{equation}
equation (\ref{equation2D}) reduces to the DLFL equation
(\ref{main}), where the signs $\sigma=\pm 1$ correspond to
$\Phi^{\ast}$ and $\Phi$ respectively.

\section*{References}


\begin{thebibliography}{59}


\bibitem{Scott_nonlin}
A.~Scott (Ed.), Encyclopedia of Nonlinear Science, Routledge, New
York,  2005.

\bibitem{Petviashvili_book1992}
O.~A. Pokhotelov, V.~I. Petviashvili, Solitary Waves in Plasmas
and in the  Atmosphere, Gordon and Breach, Reading, 1992.

\bibitem{Scorich2010}
M.~Kono, M.~M. \v{S}cori\'{c}, Nonlinear Physics of Plasmas,
Springer,  Heidelberg, 2010.

\bibitem{Kono}
M.~Kono, M.~Kawakita, Temporally and spatially pulsating solitons
in a  nonlinear stage of the long-wave Buneman instability, Phys.
Fluids B 2
  (1990) 1084-1087.

\bibitem{Lashkin_string}
V.~M. Lashkin, Blow-up solitons at the nonlinear stage of the
two-stream  instability in quantum plasmas, Europhys. Lett. 130
(2020) 30001.

\bibitem{Oikawa}
N.~Yadjima, M.~Oikawa, Formation and Interaction of Sonic-Langmuir
  Solitons: Inverse Scattering Method, Progr. Theor. Phys. 56 (1976)
  1719-1739.

\bibitem{Lashkin1991}
T.~A. Davydova, V.~M. Lashkin, Short-wavelength ion-cyclotron
soliton, Sov. J.  Plasma Phys. 17 (1991) 568-570.

\bibitem{Krall_book1973}
N.~A. Krall, A.~W. Trivelpiece, Principles of Plasma Physics,
McGraw-Hill, New   York, 1973.

\bibitem{Akhiezer_book1975}
A.~I. Akhiezer, I.~A. Akhiezer, R.~V. Polovin, A.~G. Sitenko,
K.~N. Stepanov,   Plasma Electrodynamics: Linear Theory. Vol. 1,
Pergamon, Oxford, 1975.

\bibitem{Lashkin1994}
T.~A. Davydova, A.~I. Fishchuck, V.~M. Lashkin, Short-wavelength
ion  Bernstein nonlinear waves and solitons, J. Plasma Phys. 52
(1994) 353-364.

\bibitem{Fokas1995}
A.~S. Fokas, On a class of physically important integrable
equations, Physica D
  87 (1995) 145-150.

\bibitem{Lenells2009_Nonlinearity}
J. Lenells, A. S. Fokas, On a novel integrable generalization of
the nonlinear  Schr\"{o}dinger equation, Nonlinearity 22 (2009)
11-27.

\bibitem{Kaup1978}
D.~J. Kaup, A.~C. Newell, An exact solution for a derivative
nonlinear
  Schrodinger equation, J. Math. Phys. 19 (1978) 798-801.

\bibitem{Zakharov_book}
S.~P. Novikov, S.~V. Manakov, L.~P. Pitaevski, V.~E. Zakharov,
Theory of  Solitons: The Inverse Scattering Method, Consultants
Bureau, New York, 1984.

\bibitem{Lenells2009_derivation}
J. Lenells, Exactly solvable model for nonlinear pulse propagation
in optical  fibers, Stud. Appl. Math. 123 (2009) 215-232.

\bibitem{Lenells_N-soliton2010}
J.~Lenells, Dressing for a novel integrable generalization of the
nonlinear   Schr\"{o}dinger equation, J. Nonlinear Sci. 20 (2010)
709-722.

\bibitem{Matsuno_bright2012}
Y.~Matsuno, A direct method of solution for the Fokas--Lenells
  derivative nonlinear Schr\"{o}dinger equation: I. Bright soliton
  solutions, J. Phys. A: Math. Theor. 45 (2012) 235202.

\bibitem{Veksler_dark2011}
V.~E. Vekslerchik, Lattice representation and dark solitons of the
  Fokas-Lenells equation, Nonlinearity 24 (2011) 1165-1175.

\bibitem{Matsuno_dark2012}
Y.~Matsuno, A direct method of solution for the Fokas--Lenells
  derivative nonlinear Schr\"{o}dinger equation: II. Dark soliton
  solutions, J. Phys. A: Math. Theor. 45 (2012) 475202.

\bibitem{Geng2019}
X.~Geng, J.~Shen, B.~Xue, A Hermitian symmetric space
Fokas--Lenells  equation: Solitons, breathers, rogue waves, Ann.
Phys. 404 (2019) 115-131.

\bibitem{Xu2015}
S.~Xu, J.~He, Y.~Cheng, K.~Porseizand, The $n$-order rogue waves
of  Fokas - Lenells equation, Math. Methods Appl. Sci. 38 (2015)
  1106-1126.

\bibitem{Andre1987}
M.~Andre, H.~Koskinen, G.~Gustafsson, R.~Lundin, Ion waves and
ungoing ion beams observed by Viking, Geophys. Res. Lett. 14
(1987) 463-466.

\bibitem{Cattell1991}
C.~ A. Cattell et. al., ISEE 1 observations of electrostatic ion
cyclotron waves in association with ion beams on auroral field
lines from 2.5 to 4.5 R,  J. Geophys. Res. 96 (1991) 11421-11439.

\bibitem{Temerin1997}
F.~S. Mozer, R. Ergun, M. Temerin, C. Cattel, J. Dombeck, J.
Wygant, New features of time domain electric-field structures in
the auroral acceleration region, Phys. Rev. Lett. 79 (1997)
1281-1284.

\bibitem{Lashkin2021}
V.~M. Lashkin, Perturbation theory for solitons of the
Fokas--Lenells  equation: Inverse scattering transform approach,
Phys. Rev. E 103 (2021)  042203.

\bibitem{Huang2007}
G.-Q. Zhou, N.-N. Huang, An $N$-soliton solution to the DNLS
equation based  on revised inverse scattering transform, J. Phys.
A: Math. Theor. 40 (2007)
  13607-13623.

\bibitem{Lashkin2006}
V.~M. Lashkin, Alfv\'{e}n soliton and emitted radiation in the
presence of  perturbations, Phys. Rev. E 74 (2006) 016603.

\bibitem{Vein}
R.~Vein, P.~Dale, Determinants and Their Applications in
Mathematical  Physics, Springer, New York, 1999.

\end{thebibliography}

\end{document}